\documentclass[10pt]{iopart}
\expandafter\let\csname equation*\endcsname\relax
\expandafter\let\csname endequation*\endcsname\relax
\usepackage{graphicx}
\usepackage{lipsum}
\usepackage{bm}
\usepackage{tabu}
\usepackage[utf8]{inputenc}
\usepackage[T1]{fontenc}
\usepackage{mathptmx}
\usepackage{color}
\usepackage{xcolor}
\usepackage{cite}
\usepackage{iopams}  
\usepackage{amsmath}
\usepackage{tikz}
\usepackage{bbold}
\newcommand\encircle[1]{%
  \tikz[baseline=(X.base)] 
    \node (X) [draw, shape=circle, inner sep=-.5] {\strut #1};}

\begin{document}
\title[Generation and characterization of complex vector modes with digital micromirror devices]{Generation and characterization of complex vector modes with digital micromirror devices}

\author{Xiao-Bo Hu$^1$, Carmelo Rosales-Guzm\'an$^{2,3}$}
\address{$^1$ Key Laboratory of Optical Field Manipulation of Zhejiang Province,Department of Physics, Zhejiang Sci-Tech University, Hangzhou, 310018, China.}
\address{$^2$ Centro de Investigaciones en \'Optica, A.C., Loma del Bosque 115, Colonia Lomas del campestre, C.P. 37150 Le\'on, Guanajuato, Mexico}
\address{$^3$ Wang Da-Heng Collaborative Innovation Center, Heilongjiang Provincial Key Laboratory of Quantum Manipulation and Control, Harbin University of Science and Technology, Harbin 150080, China.}
\ead{huxiaobo@hrbust.edu.cn}

\begin{abstract}
Complex vector light modes with a spatial variant polarization distribution have become topical of late, enabling the development of novel applications in numerous research fields. Key to this is the remarkable similarities they hold with quantum entangled states, which arises from the non-separability between the spatial and polarisation degrees of freedom (DoF). As such, the demand for diversification of generation methods and characterization techniques have increased dramatically. Here we put forward a comprehensive tutorial about the use of DMDs in the generation and characterization of vector modes, providing details on the implementation of techniques that fully exploits the unsurpassed advantage of Digital Micromirrors Devices (DMDs), such as their high refresh rates and polarisation independence. We start by briefly describing the operating principles of DMD and follow with a thorough explanation of some of the methods to shape arbitrary vector modes. Finally, we describe some techniques aiming at the real-time characterization of vector beams. This tutorial highlights the value of DMDs as an alternative tool for the generation and characterization of complex vector light fields, of great relevance in a wide variety of applications.
\end{abstract}
 
\ioptwocol
\maketitle
\section{Introduction}

Complex vector light fields, also known as vector or classically-entangled modes, feature a non-homogeneous polarisation distribution, resulting from the non-separable coupling between the spatial and polarisation degrees of freedom \cite{Roadmap,forbes_structured_2021,Galvez2012,toninelli2019,forbes2019classically,konrad2019}. Vector modes have drawn significant attention in the last two decades or so due to their broad variety of applications \cite{Zhan2009,Chen2014,Rosales2018Review,Yuanjietweezers2021,backscattering2021,Hu2019}. As such different techniques have been proposed for their generation, amongst which, computer-controlled devices, such as liquid crystal spatial light modulators (SLMs) and Digital micromirror devices (DMDs), stand out due to their high flexibility. SLMs, however, are polarization-dependent, allowing only the modulation of linear polarization (typically horizontal). Thus, for generating arbitrary vector modes with SLMs, the transverse profiles of both polarization components have to be manipulated independently, either in interferometric arrays containing one or two SLMs\cite{SPIEbook,Rosales2017,Maurer2007,Liu2018,Mendoza-Hernandez2019,Niziev2006,Moreno2012,Mitchell2017} or via a temporal sequence using a double pass over a single SLM\cite{Rong2014,Otte2018b}. Besides, their high price, low refresh rate (60Hz), polarization-sensitivity, wavelength dependency, and low efficiency, also represent serious disadvantages in many applications. On the contrary, recently proposed techniques based on the DMD technology, initially developed for projection systems, have shown unsurpassed advantages over SLMs, such as, high refresh rates, polarisation independence, low cost, wide operation range, amongst others. Their refresh rates for example, which can reach up to 30 kHz, provides the means to produce arbitrary vector beams fields at high speeds \cite{Hu2018,Ren2015,Chen2015DMD,Mitchell2016,Goorden2014,Lerner2012,Hu2021Random,scholes_structured_2019}. Nonetheless, in almost all the techniques proposed so far, the polarisation-insensitive attribute of DMDs has gone almost unnoticed. An exception to this is a recently proposed technique, which we will describe in detail in this tutorial \cite{rosales-guzman_polarisation-insensitive_2020}.

Another important aspect of complex light fields is their characterization, which provides useful information about their purity or their spatial polarisation distribution. Nonetheless, due to the non-separable coupling between the spatial and polarisation DoFs, their characterisation is challenging. A well-known technique, aiming at the reconstruction of their entire transverse polarisation, is Stokes polarimetry, which comprises a series of intensity measurements \cite{Goldstein2011}. A more modern approach, which exploits their similarity with quantum entangled states, utilizes well-established tools from quantum mechanics, namely, {\it Concurrence} ($C$). Such measure, which for vector modes is called Vector Quality Factor (VQF), measures the degree of coupling between the spatial and polarisation DoF assigning a number in the range $[0,1]$, 0 to scalar modes with a null degree of coupling and 1 to vector modes with a maximum degree of coupling \cite{McLaren2015,Ndagano2016,Otte2018,Bhebhe2018a,selyem2019,Ndagano2018}. This technique was first implemented with SLMs requiring the spatial separation of both polarisation components, followed by the projection of each beam onto a series of spatial filters encoded on the SLM. Noteworthy, the polarisation-insensitivity property of DMDs enables the real-time and all-digital stokes polarimetry to reconstruct the transverse polarisation distribution as well as a simplified implementation of the VQF measurement \cite{Zhao2019,manthalkar_all-digital_2020,Zhaobo2020}. As such, DMDs represent a powerful tool for the generation and characterisation of complex vector modes, of great relevance in industrial applications as well as in undergraduate optics laboratories.

In what follows, we provide a comprehensive description, in the form of a tutorial, of the use of DMDs to generate and characterise complex vector modes. First, in section \ref{generation} we briefly  describe the working principle of DMDs, followed by a tutorial-style explanation of their use in controlling the amplitude and phase of light fields (section \ref{binaryholograms}). In section \ref{Random encoding} we describe a novel binary encoding scheme based on random spatial multiplexing to maximally exploit the high refresh rates of DMDs. Afterwards, in section \ref{ExperimentalGeneration}, we describe in detail the experimental generation of complex vector modes, providing a variety of examples, such as, Laguerre-, Ince-, Mathieu- and Parabolic-Gaussian vector beams \cite{Yao-Li2020,Rosales2021,Hu2021,ZhaoBo2021}. Finally, in section  \ref{Characterzation} we focus on the description of characterization techniques, we first describe a real-time Stokes Polarimetry(SP) technique followed by the description of an alternative technique to determine the VQF. Both techniques will pave way for real-time characterisation of vector modes, of great relevance in applications where the non-separability can be used as an optical sensor. 

\section{Binary beam shaping with digital micromirror devices}
\label{generation}
\subsection{Phase and amplitude binary holograms}
\label{binaryholograms}
\begin{figure}[tb]
   \centering
    \includegraphics[width=0.49\textwidth]{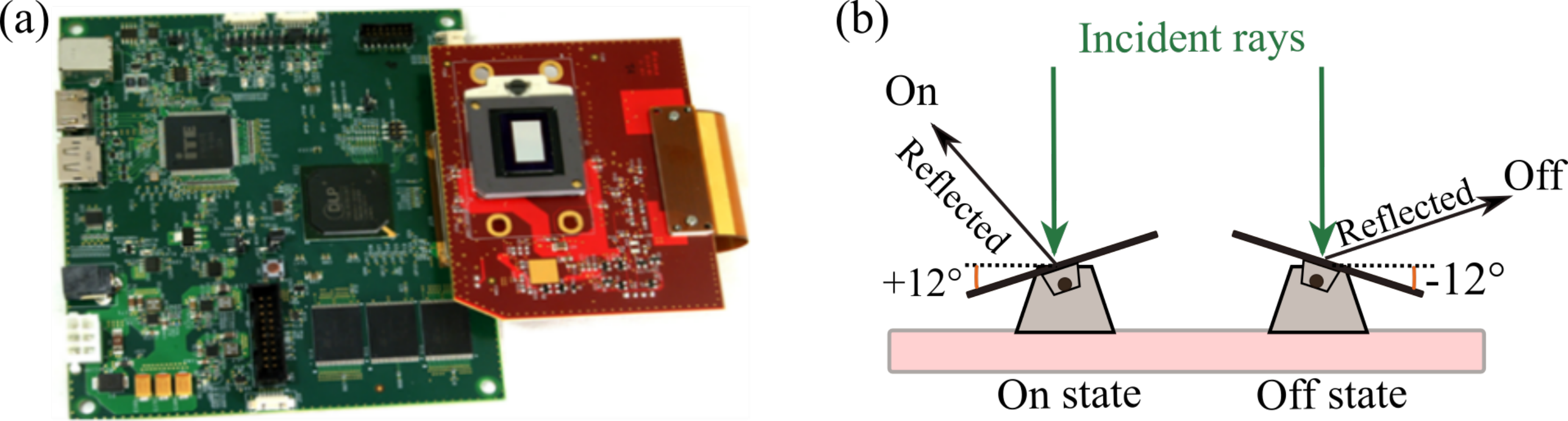}
    \caption{(a) Digital micromirror device (DLP Light Crafter 6500 from Texas Instruments) and (b) its schematic diagram.}
   \label{dmd}
\end{figure}

DMDs were initially designed as digital projection devices but nowadays are routinely used as spatial light modulators. Figure \ref{dmd}(a) shows an example of the specific DMD (DLP Light Crafter 6500 from Texas Instrument), which was employed in all the DMD techniques introduced across the whole tutorial. A DMD consist of an array of millions of micro-sized mirrors ($\approx 8 \mu$m in size), each of which can be turned to an "Off" or "On" state by tilting it $-12^\circ$ or $+12^\circ$, respectively. In this way, when the DMD is properly aligned, each mirror in the "On" state reflects light in the desired direction\cite{scholes_structured_2019,Cox2021}, as schematically illustrated in Fig. \ref{dmd}(b). Since each mirror can only be in two states, it is important to bear in mind that one of the principal requirements is to address the DMD with binary holograms. Nonetheless, binary-amplitude holograms are capable to shape both the amplitude and phase of a complex light field. It is worth mentioning that even some devices can also display gray-scale holograms by allowing the micromirrors to oscillate between the "On" or "Off" states, their refresh rates are dramatically reduced.

\begin{figure}[tb]
   \centering
    \includegraphics[width=0.45\textwidth]{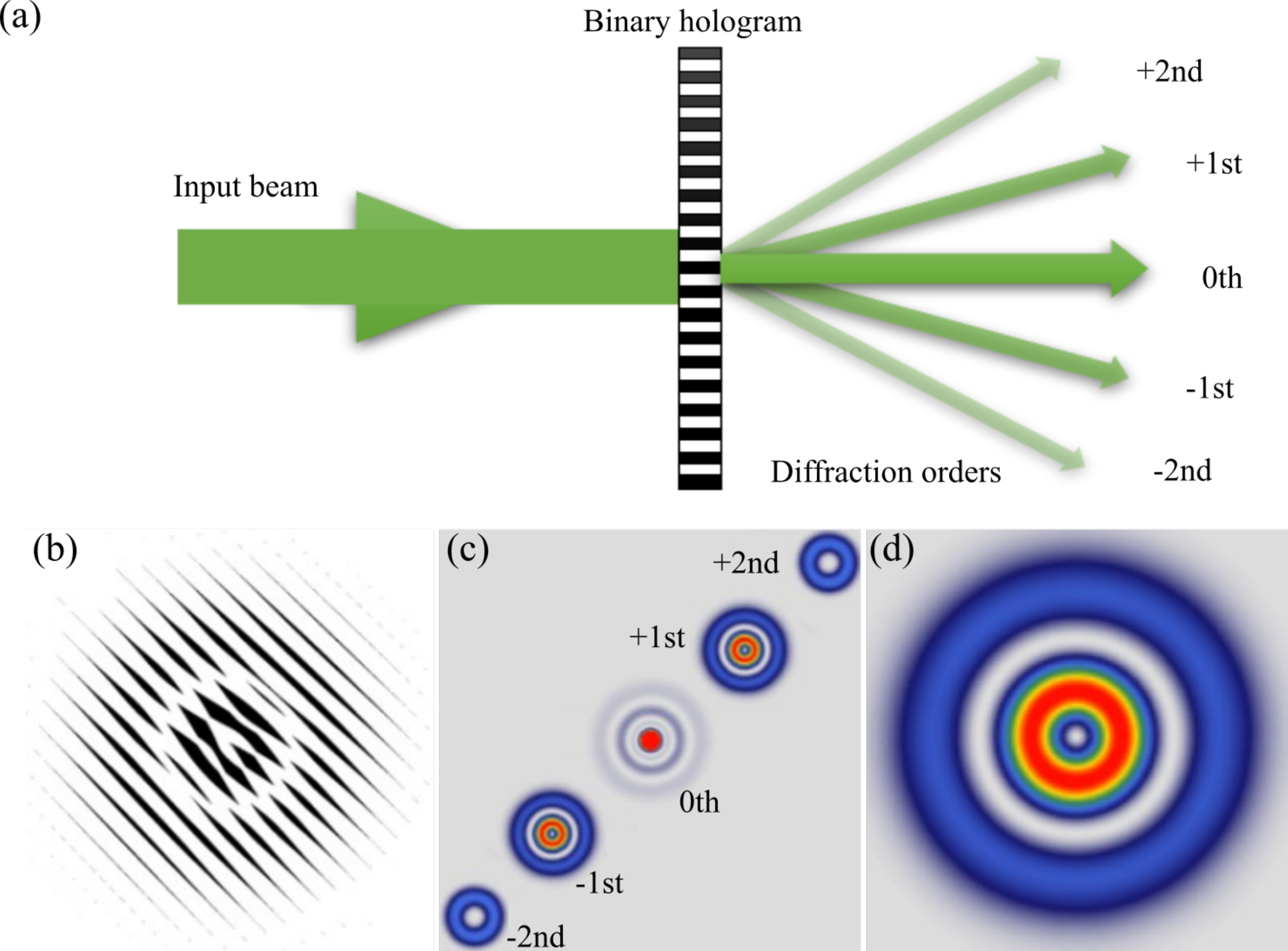}
    \caption{(a) A binary periodic grating diffracts the expanded input beam into multiple orders. (b) Example of a binary hologram showing the mode $LG_{1}^{-1}$. (c) Diffraction patterns caused  when such hologram is illuminated by a flat wavefront, only the first five diffraction orders are shown. (d) The first diffraction order.}
   \label{DifOrders}
\end{figure}

The binary transmittance function $T(x,y)$ of a typical binary hologram designed to engineer a scalar complex light field $u(x,y)=A(x,y)\phi(x,y)$, with amplitude $A(x,y)$ and phase $\phi(x,y)$ is expressed as \cite{Mitchell2016},

\begin{equation}
T(x,y)=\frac{1}{2}+\frac{1}{2}\text{sgn}\left\{\cos\left[p(x,y)\right]+\cos\left[q(x,y)\right]\right\},
\label{TM}
\end{equation}
where the amplitude and phase information are given by,
\begin{equation}
\begin{split}
q(x,y)&=\arcsin\left({A(x,y)}/{A_{ max}}\right),\\
p(x,y)&=\phi(x,y)+2\pi(\nu x+\eta y),
\end{split}
\end{equation}
respectively. Here, sgn$\{\cdot\}$ is the sign function that reflects the binary-amplitude modulation, which forces all arguments to fall to either 0 or 1. The term ${A_{max}}$ represents the maximum amplitude value and the phase term $2\pi(\nu x+\eta y)$ is an additional linear diffraction grating with spatial frequency  ($\nu, \eta$), responsible for the generation of multiple diffraction orders. Noteworthy, the desired mode $u(x,y)$ is generated in the first diffraction order. In addition, the desired amplitude $A(x,y)$ and phase exp$(i\phi(x,y))$ are modulated by locally varying the width and lateral position, respectively, of each diffraction grating. Importantly, the angle of the first diffraction order, and therefore its position in the observation plane can be adjusted via the spatial frequency of the grating. Figure \ref{DifOrders}(a) conceptually represents such binary hologram where a binary periodic grating diffracts the input beam into multiple orders. A typical example of a binary hologram for generating the specific Laguerre-Gaussian mode ($LG_{p}^{\ell}$) with radial index $p=1$ and azimuthal index $\ell=-1$ is shown in Fig. \ref{DifOrders}(b). Notice that the colors in all the binary holograms shown through the manuscript were inverted and the holograms ware also scaled for display purposes. Hence, the black color represents the mirrors in the "On" state that reflect the light in the desired direction, and the white color represents mirrors in the ‘Off’ state, which direct light away. In the actual experimental implementation, the negative of such transmission grating is displayed on the DMD. When this fork-like hologram is illuminated with an expanded light beam, several diffraction orders appear in the far field, five of which are shown in  Fig. \ref{DifOrders}(c). The target beam appears in the first diffraction order, which for the sake of clarity is shown in Fig. \ref{DifOrders}(d). So far, we have explained how to generate arbitrary scalar modes characterised by a homogeneous transverse polarisation distributions, in the following section we will explain how to generate complex vector light fields bearing non-homogeneous transverse polarisation distributions.

\subsection{Spatial random multiplexing encoding}
\label{Random encoding}

\begin{figure}[tb]
   \centering
    \includegraphics[width=0.49\textwidth]{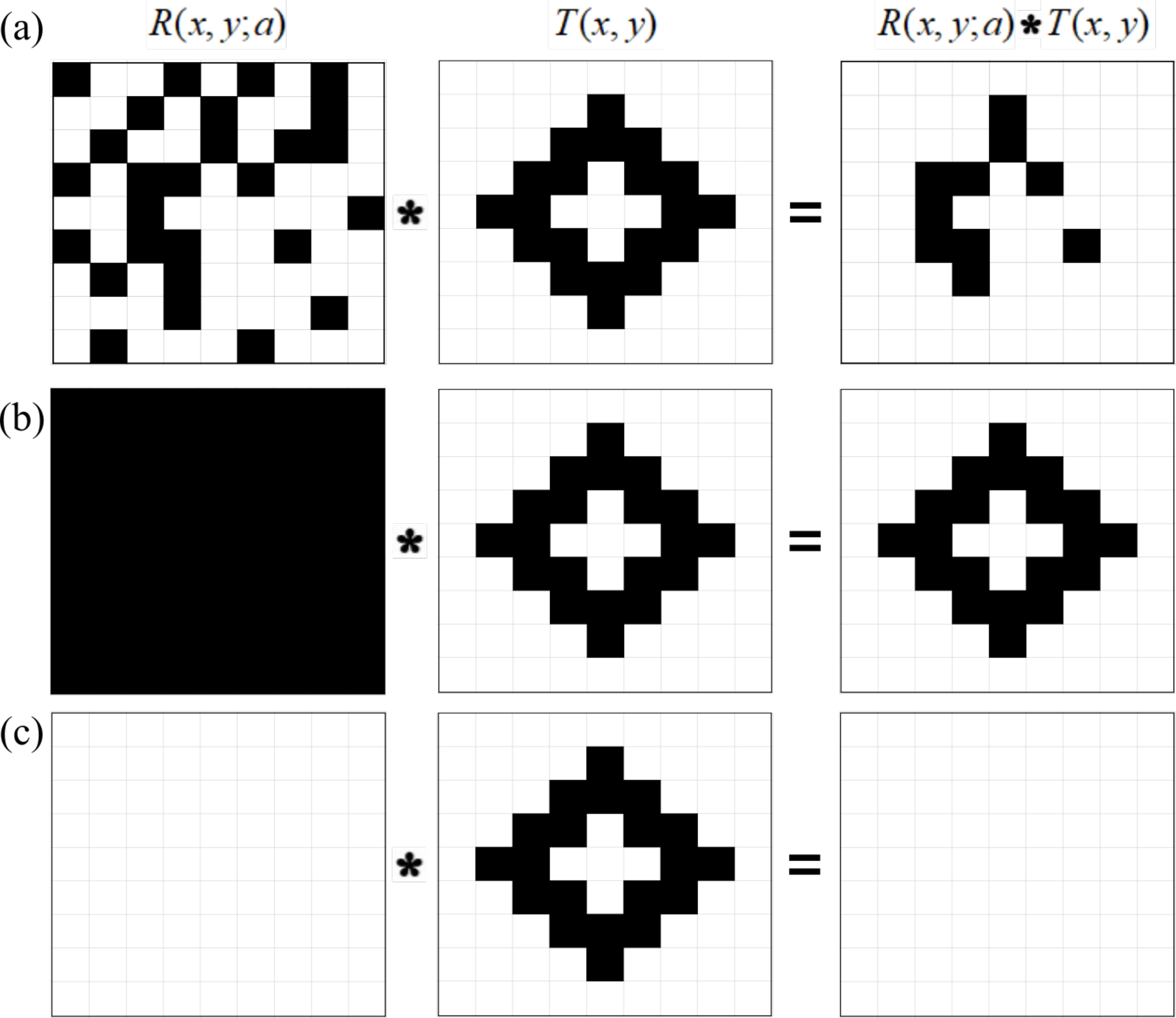}
    \caption{The convolution of a random binary mask $R(x,y;a)$ (left) and a transmittance function $T(x,y)$ (middle) give rise to the synthetic mask shown on the right. Here, we show the specific cases for $a\in [0.33,1,0.06]$ in (a), (b) and (c), respectively. The pixels staying in the "Off" state represented with the color of white will direct the light away while the ones with black color staying in the "On" state will redirect the light to the desire direction. Notice how the pixels with black color in the random binary mask $R(x,y;a)$ switch to after convoluted with the transmittance function $T(x,y)$.}
   \label{Sparandom}
\end{figure}

On the basis of the device introduced earlier, an encoding scheme of spatial random multiplexing, specially designed for the DMD working mechanism is introduced in this section. Crucially, such scheme allows to maximally exploit the high refresh rates of DMDs during the generation of complex vector modes. In this encoding scheme, we firstly define a random binary mask $R(x,y;a)$  with the exact resolution of the DMD (1920$\times$1080 pixels). Examples of such mask containing 9$\times$9 pixels are schematically shown in the left panels of Fig.\ref{Sparandom}. The pixels in the "On" state, which are represented with the black color, are spatially selected at random, the number of which is selected through the parameter $a\in [0, 1]$. For example, the case $a=0.5$, with half of the total pixels in the "On" state, is shown in the left panel of Fig. \ref{Sparandom}(a), whereas the case $a=1$, with all the pixels in the "On" state, is shown in the left panel of Fig. \ref{Sparandom}(b). Finally the case $a=0$, with all the pixels in the "Off" state, is illustrated in the left panel of Fig. \ref{Sparandom}(c). Thus, when such a random binary mask is displayed on the DMD, the effect of tilting the micromirrors to the ‘On’ or the ‘Off’ state is determined by the parameter $a$. Given that only the pixels in the "On" state diffract light in the desired direction, the parameter $a$ controls the amount of power in the generated mode. Notice that the power efficiency of any binary hologram is naturally limited to a maximum of $\approx 10 \%$, which in our technique is achieved when $a=1$. Now, to generate a spatial mode given by a transmittance function $T(x,y)$ (Eq. \ref{TM}), the random binary mask $R(x,y;a)$ is convoluted with this transmittance function (see middle panels of Fig.\ref{Sparandom}). As result of such convolution, only the pixels which are in the ‘On’ state on both, $T(x,y)$ and $R(x,y;a)$ will remain in the "On" state, the rest will be switched "Off", as schematically illustrated in the right panels of Fig. \ref{Sparandom}.

We are now ready to explain how the principle of random spatial multiplexing can be applied to the simultaneous generation of two (or more) optical fields with independent amplitudes and phases\cite{RosalesGuzman2013Airy,MartinezFuentes2018,Hu2021Random}. At this stage, it is worth clarifying that each optical field must be associated to an independent transmittance function, namely, $T_1(x,y)$ and $T_2(x,y)$. To generate the final transmittance function $T_{f}(x,y)$, both transmittance function are carefully recombined ensuring the removal of any spatial overlap between the pixels associated to $T_1(x,y)$ and those associated to $T_2(x,y)$, here is where the random binary approach becomes relevant. To this end, we first define the complementary function $\Bar{R}(x,y;a)$ as,
\begin{equation}
  \Bar{R}(x,y;a)=\mathbb{1}-R(x,y;1-a)
\end{equation}
where $\mathbb{1}$ represents a binary mask with all its entries in 1, {\it i.e.}, all the micromirrors in the "On" State. Importantly, the direct product, entry-by-entry, of the matrix $R(x,y;a)$ and its complementary matrix yields a mask with all its entries in zero, that is, $R(x,y;a)\Bar{R}(x,y;1-a)=\mathbb{0}$. To further clarify this, for $a=1$ all the entries of $R(x,y;1)$  are in the ‘On’ state, whereas in $\Bar{R}(x,y;0)$ all are in the ‘Off’ state. On the contrary, for $a=0$, all the pixels of $R(x,y;0)$ are in the ‘Off’ state, while in $\Bar{R}(x,y;1)$ , all are in the ‘On’ state. Using this two random binary masks, the multiplexed binary mask is then defined as,
\begin{equation}
T_{f}(x,y)=R(x,y;a)*T_1(x,y)+\Bar{R}(x,y;1-a)*T_2(x,y),
\label{TMultiplexed}
\end{equation}
\noindent

\begin{figure*}[tb]
   \centering
    \includegraphics[width=0.97\textwidth]{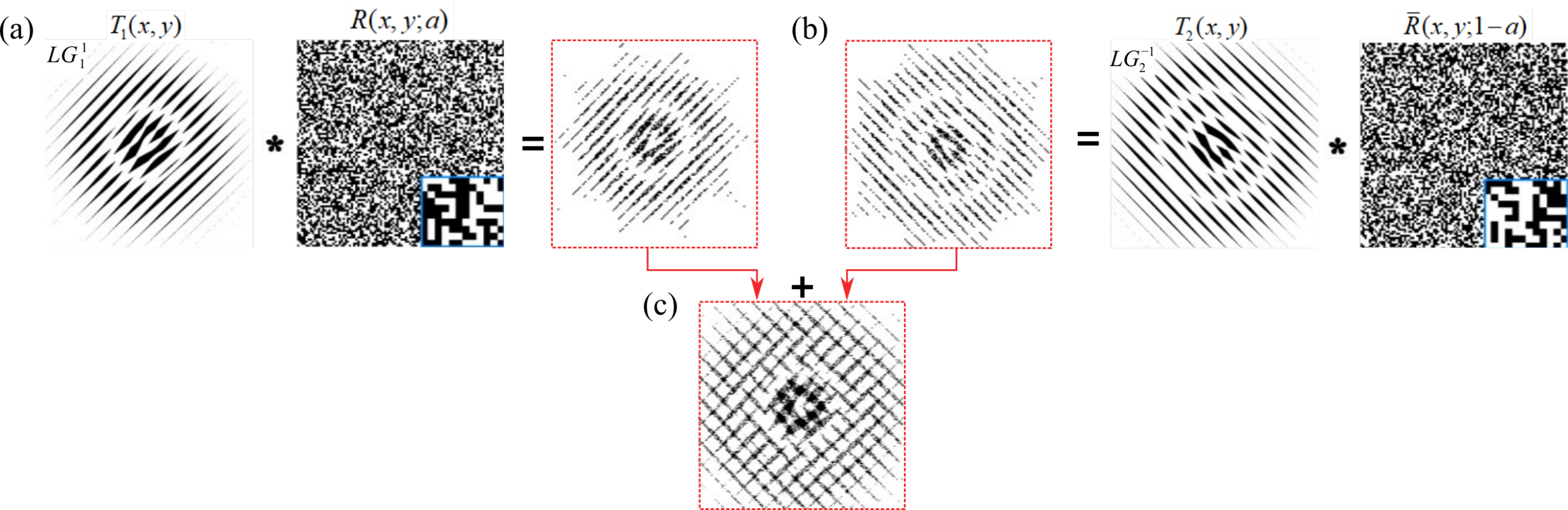}
    \caption{Schematic representation of the spatial random multiplexing encoding principle. The holograms in red dash box shown in (a) and (b) show the convolution $T_1(x,y)*R(x,y;a)$ and $T_2(x,y)*\Bar{R}(x,y;1-a)$, where the random binary mask $\Bar{R}(x,y;1-a)$ is complement of $R(x,y;a)$, as can be clearly seen in the enlarged portion of the small inset in each. $T_1(x,y)$ (left panel in (a)) and $T_2(x,y)$(middle panel in (b)) are two examples of the transmittance function which are encoded with the modes $LG_2^{-1}$ and $LG_1^{1}$. The resultant multiplexing hologram composed of (a) and (b) is shown in (c). }
   \label{Holo}
\end{figure*}

This synthetic hologram allows the simultaneous generation of two modes, with a power determined by the parameter $a$. For example, for $a=0.5$, half of the pixels, spatially selected at random, are used to generate the mode encoded in $T_1(x,y)$, while the rest to generate the mode encoded in $T_2(x,y)$. As mentioned earlier, none of the pixels associated to $T_1(x,y)$ overlap spatially with those associated to $T_2(x,y)$. Fig.\ref{Holo} schematically illustrate this description, where we encode the specific mode $LG_1^{1}$ in the function $T_1(x,y)$ and  $LG_2^{-1}$ in $T_2(x,y)$, as shown in the left panel of Fig. \ref{Holo} (a) and the middle panel of Fig. \ref{Holo} (b), respectively. We further show an example of a random binary mask $R(x,y;a)$  and its complementary random mask  $\Bar{R}(x,y;1-a)$ in the middle panel of  Fig.\ref{Holo}(a) and right panel of Fig.\ref{Holo}(b), an enlarged portion of both masks is shown as an inset to emphasize their complementary relation. As a result, the holograms of the convolution of $R(x,y;a)*T_1(x,y)$ and $R(x,y;1-a)*T_2(x,y)$ are shown in the right and left panel of Fig.\ref{Holo}(a) and Fig.\ref{Holo}(b). Finally, the target binary hologram resulting from the superposition of both holograms is shown in Fig.\ref{Holo}(c). Importantly, such a multiplexing hologram enables to generate arbitrary vector mode at the speed only limited by the specific of DMDs.

\begin{figure}[tb]
   \centering
    \includegraphics[width=0.49\textwidth]{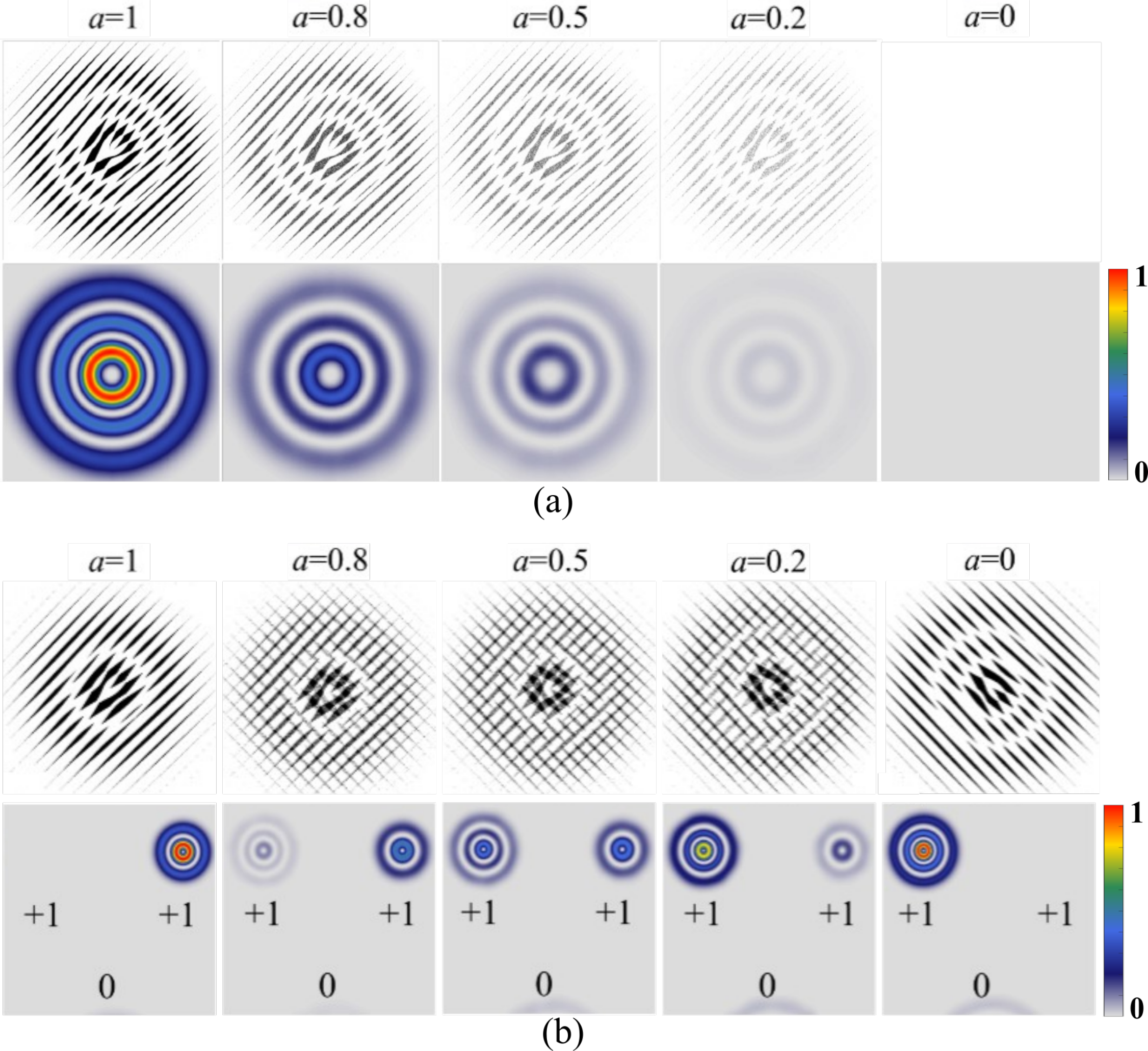}
    \caption{(a) The holograms (top) for the target mode $LG_2^{2}$ and the intensity profiles of its corresponding generated beam (bottom) varies as a function of parameter $a$, where the cases for $a=1,0.8,0.5,0.2$ and $a=0$ are shown from left to right.  Similarly, an example of a multiplexed mask for both modes $LG_1^{1}$ and  $LG_2^{-1}$ are represented in top of (b), notice how the associated intensity profile (bottom) varies as a function of $a$.}
   \label{Afunction}
\end{figure}

As stated earlier, the intensity distribution of each generated beam varies as a function of parameter $a$. To further explain this, the top row of Fig. \ref{Afunction}(a) shows the binary holograms obtained for five different values of $a$, for the target mode $LG_2^{2}$. The intensity distribution of the modes generated with such holograms, in the first diffraction order, is illustrated in the bottom row. Notice how the power of the mode decreases as the number of pixels in the "On" state also decrease. Similarly, the top row of Fig. \ref{Afunction}(b) shows the same effect for a multiplexed mask with a transmittance function corresponding to the modes $LG_1^{1}$ and $LG_2^{-1}$. The transmittance function corresponding to each mode was generated with different spatial frequencies, for this example both have the same frequency in the vertical direction ($\eta_1=\eta_2$), but differ in the horizontal direction ($\nu_1=-\nu_2$). In this case we obtain two modes in the $1^{st}$ diffraction order, which are shown in the bottom row of the same figure. Notice that, for  $a=1$, only the hologram corresponding to the $LG_1^{1}$ mode is present, as $a$ decreases, it gradually disappears until it vanishes entirely for $a=0$, while the contrary happens for the hologram corresponding to the mode $LG_2^{-1}$. In regards to the intensity distribution, the power of the $LG_1^{1}$ mode decreases gradually while the power of the $LG_2^{-1}$ mode increases. Importantly, for the specific case $a=0.5$, both beams are present.
\section{Experimental generation of complex vector beams}
\label{ExperimentalGeneration}
\subsection{Theoretical description}
As it is well-known, in vector light fields the spatial and polarization DoFs are coupled in a non-separable way, which can be expressed mathematically as \cite{Galvez2012,Chen2014},
\begin{equation}
\centering
\vec{u}(\vec{r}) =  \cos \left(\theta\right) u_{1}(\vec{r})\hat{r}+ \sin \left(\theta\right) u_{2}(\vec{r}){\text e}^{i\alpha} \hat{l},
\label{Vectormodes}
\end{equation}
\noindent
where $u_{i}(\vec{r})$ represent a pair of orthogonal spatial modes. Further, $\hat{r}=(\hat{h}+i \hat{v})/\sqrt{2}$ and $\hat{l}=(\hat{h}-i \hat{v})/\sqrt{2}$ represent the unitary vectors of the right- and left-handed circular polarisation, while  $\hat{h}$ and $\hat{v}$ represent the unitary vectors of the horizontal and vertical polarisation basis, respectively. Noteworthy, while we use the circular basis ($\hat{r},\hat{l}$), any other orthogonal polarisation basis can also be used, linear ($\hat{h},\hat{v}$), diagonal ($\hat{d},\hat{a}$), or elliptical ($\hat{er},\hat{el}$). The coefficients $\cos(\theta)$ and $\sin(\theta)$ ($\theta\in[0,\pi/2]$) are weighting factors that allow a smooth transition of the field $\vec{u}(\vec{r})$, from scalar ($\theta=0$ and $\theta=\pi/2$) to purely vector ($\theta=\pi/4$). Additionally, the term ${\text e}^{i\alpha}$ ($\alpha\in[-\pi/2,\pi/2]$) adds an extra phase delay between both polarisation components.

It is worth mentioning that the spatial degree of freedom encoded in the spatial modes $u_1(\vec{r})$ and $u_2(\vec{r})$ can be given by any of the solutions to the wave equation in the different coordinates systems. An example of such are the Laguerre-Gaussian (LG) vector modes, which are natural solutions in cylindrical coordinates. From Eq. \ref{Vectormodes} the LG vector modes can be expressed as,
\begin{equation}
\centering
\vec{u}(\vec{r}) = \cos \left(\theta\right) LG_{p_1}^{\ell_1}(\vec{r})\hat{r}+ \sin \left(\theta\right) LG_{p_2}^{\ell_2}(\vec{r}){\text e}^{i\alpha} \hat{l},
\label{VectormodesLG}
\end{equation}
\noindent 
where $\vec{r}=(\rho,\phi)$ is the position vector of the cylindrical coordinates $(\rho,\phi)$ and $LG_{p}^{\ell}$ are the Laguerre-Gaussian modes, carrying a well-defined amount of Orbital Angular Momentum (OAM), as defined in \cite{Hu2020}.

\begin{figure*}[tb]
   \centering
    \includegraphics[width=0.95\textwidth]{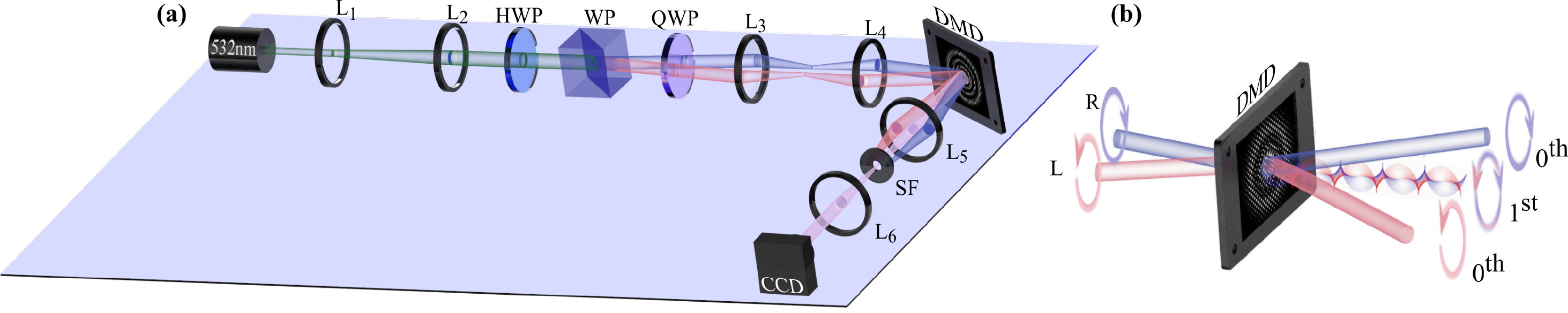}
    \caption{ (a) Schematic representation of the experimental setup for the polarization-insensitive generation of arbitrary vector mode based on DMDs. (b) The concept of how the linear diffraction grating effect the angle of the first diffraction order. Here, the multiplexing hologram encoded on the DMD is the superposition of two independent holograms with unique spatial carrier frequencies. Such parameters need to be carefully picked to guarantee the overlap of the first diffraction order from each beam along the same propagation path.}
   \label{Setupge}
\end{figure*}
\subsection{Experimental setup}
Complex vector modes of various spatial distributions can be generated from a DMD in different ways. Here, we will describe an approach that takes full advantage of the polarisation-insensitive property of DMDs, which is fully detailed in  \cite{rosales-guzman_polarisation-insensitive_2020}. The experimental setup for its implementation is schematically shown in Fig. \ref{Setupge}(a). To begin with, a horizontally polarized laser beam ($\lambda=523$ nm) is collimated  and expanded by lenses $L_1$ ($f_1=20$mm) and $L_2$ ($f_2=200$mm). Afterwards, the polarisation state of the expanded beam is rotated to the diagonal polarisation state with the help of a Half-Wave Plate (HWP) at $22.5^\circ$. A Wollaston Prism (WP) subsequently separates the beam into its horizontal and vertical polarization components, both of which are transformed to the circular polarization basis ($\hat{l}, \hat{r}$) after passing a Quarter-Wave Plate (QWP). A $4f$ imaging system composed of lenses $L_3$ and $L_4$ ($f_3=f_4=200$mm) redirects these two beams towards the center of a DMD where they impinge at slightly different angles ($\approx 1.5^\circ$ ) but exactly at the same spatial location, the center of the hologram. A multiplexed binary amplitude hologram based on the spatial random multiplexing method, which will be explained in the following section, is encoded on the DMD. Such multiplex hologram consist of the superposition of two individual holograms, one for each of the constituting scalar fields in Eq. \ref{Vectormodes}, overlapped with a controllable linear diffraction grating. The period of the diffraction grating is carefully chosen to ensure the overlap of the first diffraction order of each beam along a common propagation path, where the desired complex vector field  $\vec{u}(\vec{r})$ is generated.  

We can explain this process in a more detailed way using the schematic representation shown in Fig. \ref{Setupge}(b). Here and for the sake of clarity the DMD is represented as a transmission device but as shown in Fig. \ref{Setupge}(a) it is a reflection device. The two input beams with orthogonal circular polarization $\hat{r},\hat{l}$, which emerge after the QWP, impinge in the center of hologram displayed on DMD. Afterwards, each beam's 0th diffraction order propagate in a divergent way from each other, as if they were simply reflected from a mirror, leaving both 1st orders propagating along the same axis where the desired beam is acquired. During the whole experimental alignment, the period of the linear diffraction grating is carefully adjusted to guarantee the overlap of both beams. Finally, to remove all the undesired diffraction orders, a spatial filter (SF) is placed in the far-field plane of a telescope composed of the lenses $L_5$ and $L_6$ ($f_5=f_6=100$mm). For the sake of clarify, in Figs. \ref{Setupge}(a) and \ref{Setupge}(b) only the first and zero diffracting orders are shown. Noteworthy, once the device is properly aligned, it enables the generation of arbitrary vector modes with almost any spatial distribution, as we will show in the next section \cite{Yao-Li2020,Hu2021,Rosales2021,ZhaoBo2021}. 

\subsection{Experimentally generated vector modes}

\begin{figure}[ht!]
   \centering
    \includegraphics[width=0.49\textwidth]{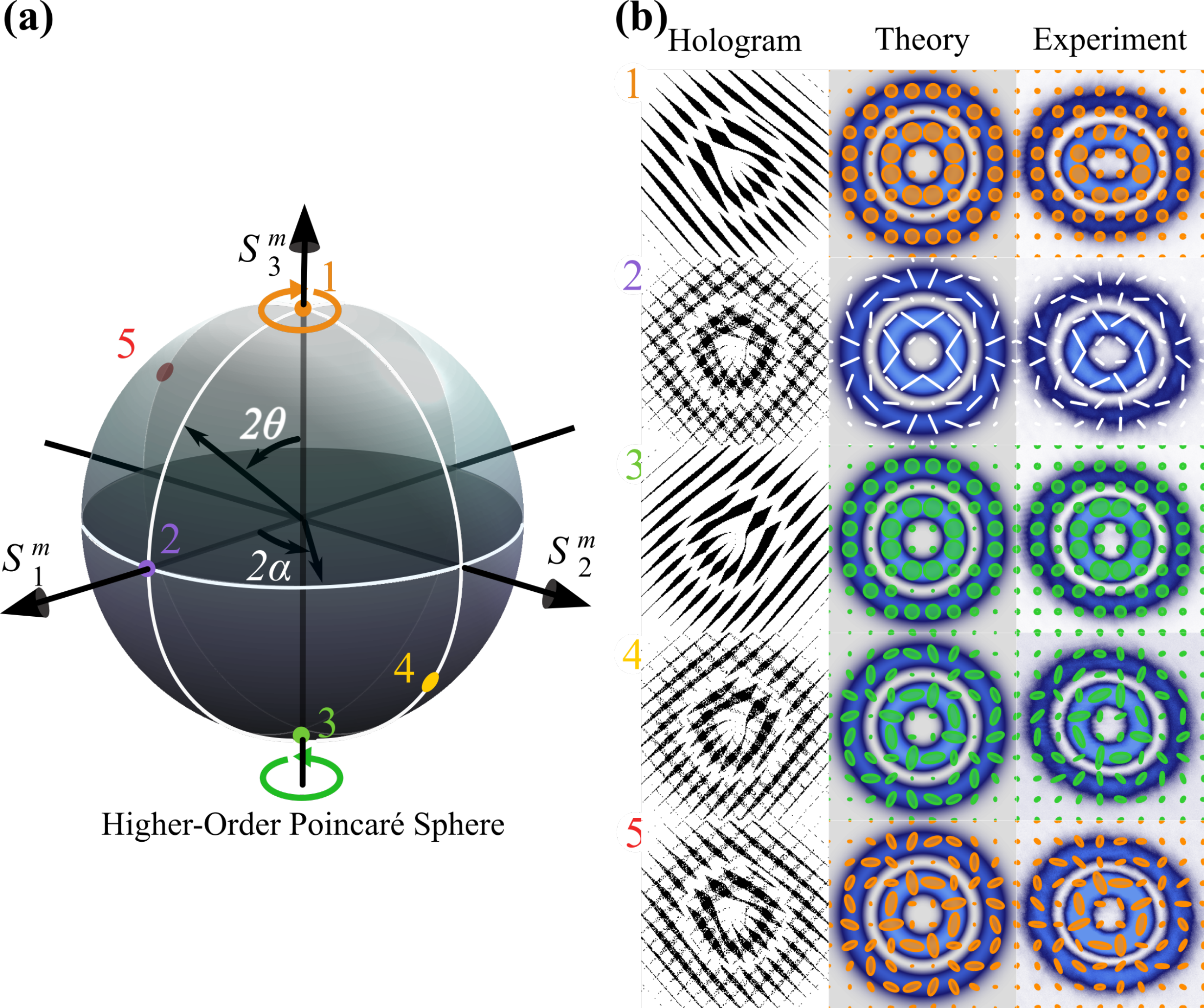}
    \caption{(a) Geometric representation of generated cylindrical vector modes on the HOPS. The required holograms to generate the specific modes labeled from 1 to 5 are shown on the left column of (b). The middle column represents the simulation results of the polarization distribution overlapped with the intensity profiles while the right column showing the corresponding experimental results.}
   \label{PoincareLG}
\end{figure}

The generation of arbitrary vector modes will be exemplified using the set of Laguerre-Gaussian vector modes and their represention on the Higher-Order Poincar\'e Sphere (HOPS). The HOPS provides a geometric way to visualize any vector mode on its surface by associating their inter-modal phase ($\alpha$) and weighting coefficient ($\theta$) to a point with coordinates ($2\alpha, 2\theta$) \cite{Milione2011}. A representative set of five $LG_p^{\ell}$ vector modes labeled from 1 to 5 and with specific coordinates (0,0), ($\pi/2$,0), ($\pi$,0), ($\pi/2$,$3\pi/4$) and ($3\pi/2$,$\pi/4$) on the HOPS, as represented on Fig.\ref{PoincareLG}(a), are shown of Fig \ref{PoincareLG}(b). Here, the left column illustrates in a conceptual way and to scale the holograms required to generate these modes, additionally, the middle and right columns shows numerical simulation and experimental results, respectively, of the transverse intensity pattern overlapped with polarization distribution. For this example we used the spatial modes $LG_1^{3}$ and $LG_1^{-3}$.

\begin{figure}[tb]
   \centering
    \includegraphics[width=0.49\textwidth]{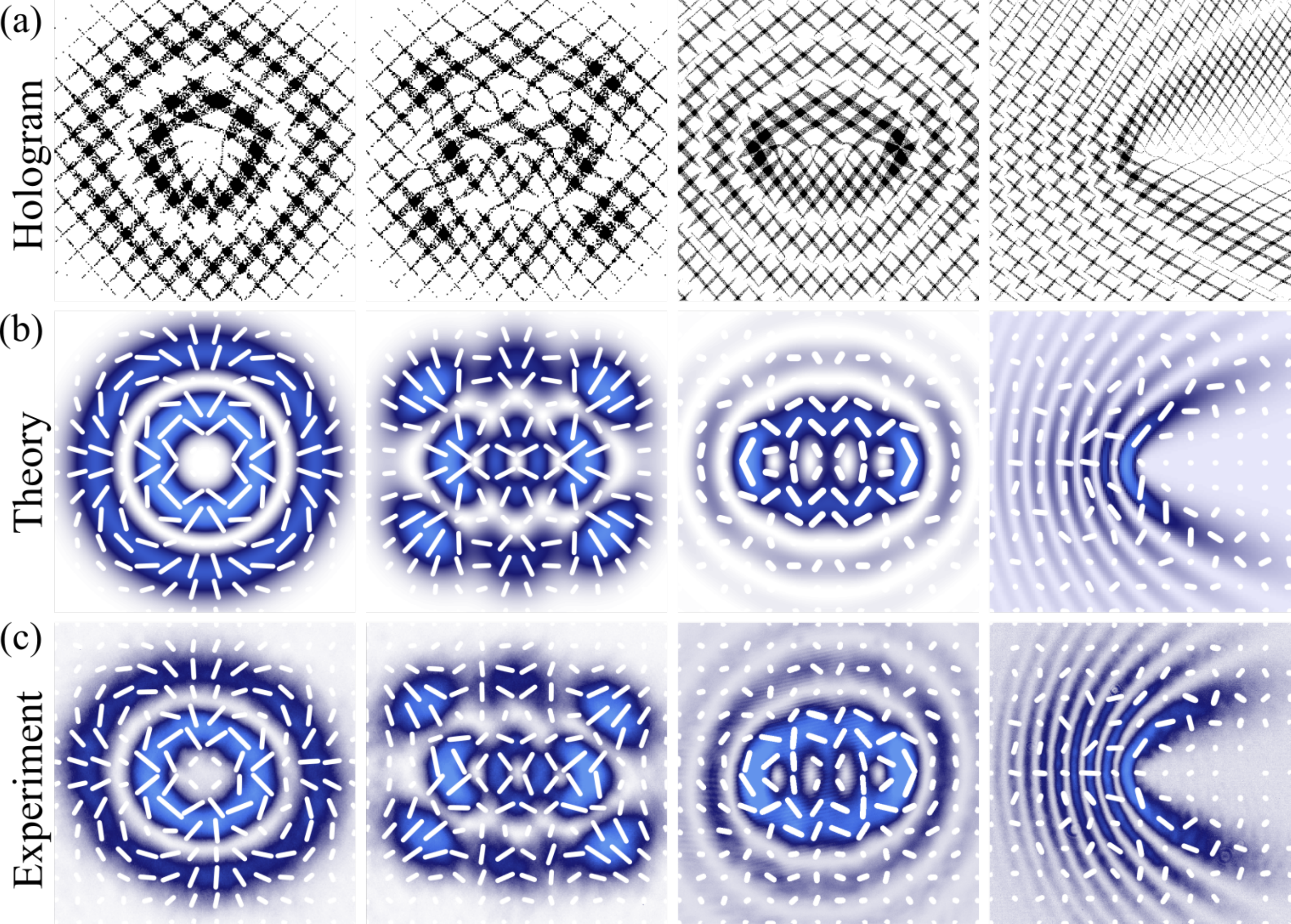}
    \caption{Representation of various vector modes. (a) From left to right holograms to generate cylindrical, Ince-, Mathieu- and Parabolic-Gauss vector modes. (b) and (c) show the theoretical and experimental intensity and transverse polarisation distribution, respectively.}
   \label{Stokesthree}
\end{figure}

Additional examples of vector beams in different coordinate systems are shown in Fig. \ref{Stokesthree}, from left to right they correspond to the Laguerre-, Ince-, Mathieu-, and Parabolic-Gaussian vector modes \cite{Yao-Li2020,Rosales2021,Hu2021}. The first is a solution to the wave equation in cylindrical coordinates, the following two are solutions in elliptical coordinates and the last is a solution in the parabolic coordinates. In Fig. \ref{Stokesthree}(a) we show, to scale, the required holograms to generate the vector modes shown in Fig. \ref{Stokesthree}(b) and \ref{Stokesthree}(c), theory and experimental, respectively, where the polarization distribution is overlapped with the transverse intesnity profile.

\section{Characterization of vector modes using a digital micromirror device}
\label{Characterzation}

Even though recent years have witnessed a growing interest in the generation of vector light fields in a variety of ways, only a few techniques have been proposed to characterise them. Along this line, DMDs have also started to play a fundamental role, as such in this section we will present two techniques capable to monitor in real time their transverse polarisation distribution and their vector quality, respectively.

\subsection{Real-time polarisation reconstruction  through  Stokes polarimetry}
Stokes polarimetry is a powerful technique that allows to reconstruct the transverse polarisation distribution of any light field through a minimum of four intensity measurements. The information contained in these intensities is captured by a set of four parameters known as Stokes parameters, from which the polarisation distribution can be determined \cite{Goldstein2011}. The relation between the intensities and the Stokes parameters is given as
\begin{equation}\label{Eq.SimplyStokes}
\begin{split}
\centering
 &S_{0}=I_{0},\hspace{19mm} S_{1}=2I_{H}-S_{0},\hspace{1mm}\\
 &S_{2}=2I_{D}-S_{0},\hspace{10mm} S_{3}=2I_{R}-S_{0},
\end{split}
\end{equation}
where $I_{0}$ is the total intensity of the given optical field. $I_{H}$ and $I_{D}$ are the measured intensities of the field after a linear polarized orientated at $0$ and $\pi/4$, respectively. $I_{R}$ is the intensity acquired after the combination of a QWP at $ \pi/4$ and a linear polariser at $\pi/2$.
\begin{figure}[ht!]
   \centering
    \includegraphics[width=0.49\textwidth]{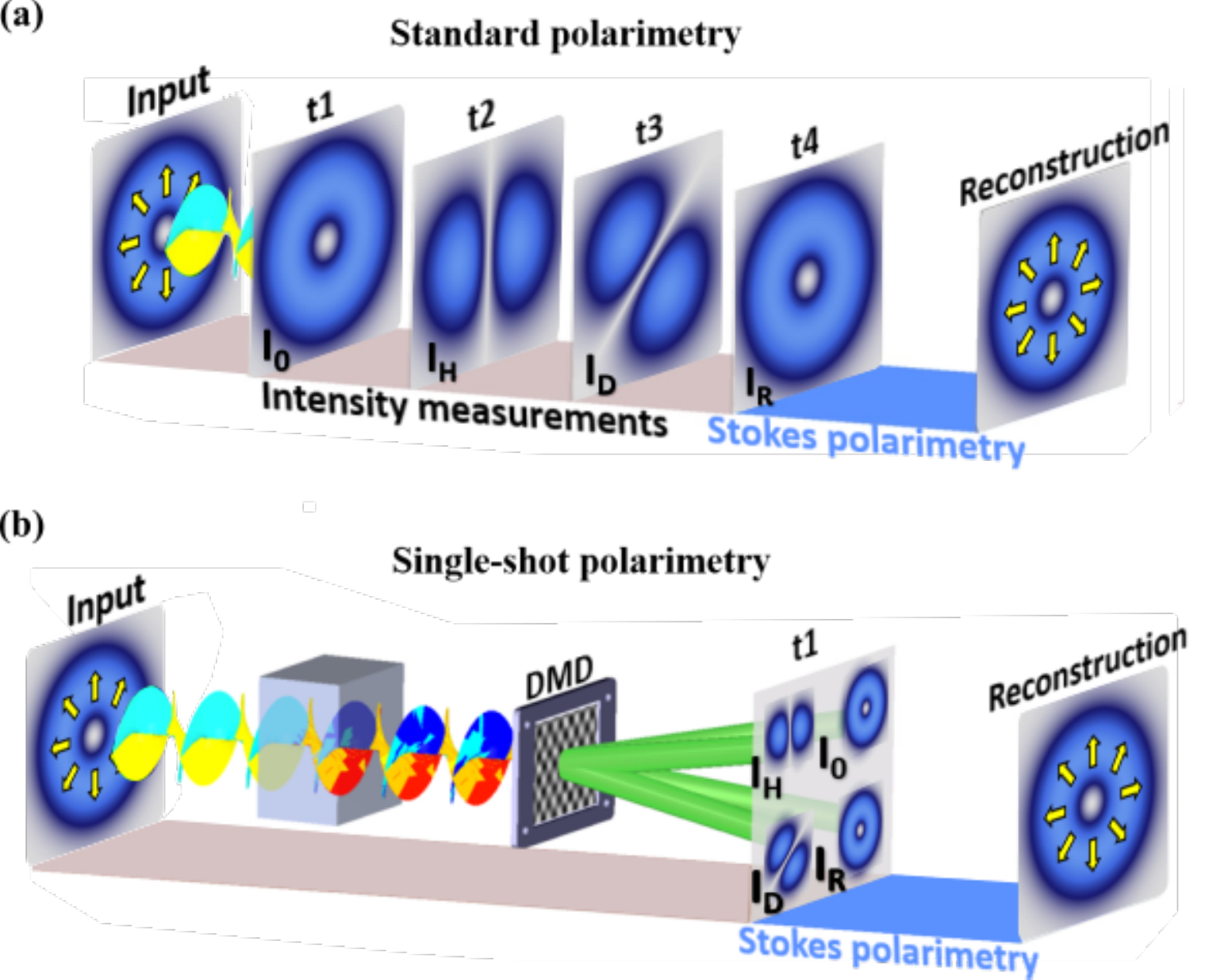}
    \caption{(a) The required intensities to reconstruct the polarisation distribution of any light field in standard polarimetry are commonly recorded one by one at different time $t_1$, $t_2$, $t_3$ and $t_4$. (b) Our technique enables a simultaneous measurement of the intensities by taking advantage of the polarisation-insensitive DMDs, which allows the splitting of the input beams into four identical copies.}
   \label{realtimeconcept}
\end{figure}

\begin{figure}[h!]
   \centering
    \includegraphics[width=0.4\textwidth]{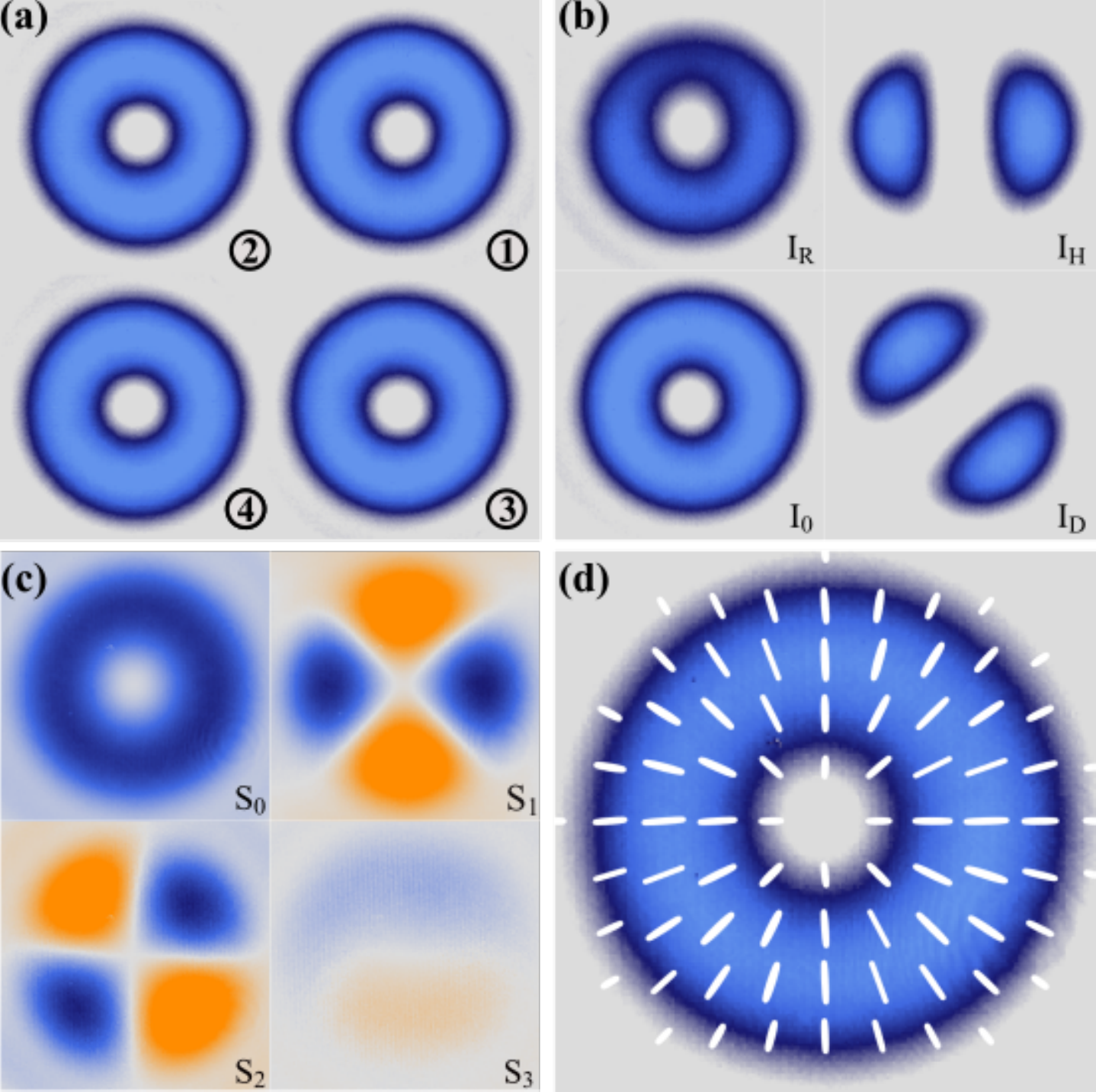}
    \caption{(a) Single-shot of the calibration image to find the centers of the beams. (b) An example of the intensity images of $I_0$, $I_H$, $I_D$ and $I_R$. (c) A set of four Stokes parameters computed through four intensity measurements from (b) simultaneously. (d) Reconstructed polarisation distribution from the Stokes parameters in (c).}
   \label{resultrealtime}
\end{figure}

Traditionally, these intensities are obtained individually, one-by-one, at different times, as schematically shown in Fig. \ref{realtimeconcept}(a)  limiting its performance to light beams with static states of polarisation. As such, in recent time we proposed  a technique relying on a DMD that allows the real-time dynamic reconstruction of the SoP of any light field, which is described next. The key idea behind this novel technique relies on performing all the intensity measurements simultaneously. For this  we exploit the polarisation-insensitive attribute of DMDs. To be more specific, a DMD is addressed with a multiplexed digital hologram, enabling to split the input beam into four identical copies propagating along different paths. In this way, all the four required intensities can be recorded simultaneously in a single shot with the help of the required optical filters and a CCD, as illustrated in Fig.\ref{realtimeconcept}(b). A dedicated software specifically designed to analyse the acquired images allows to reconstruct in real time the state of polarization of any light beam at speeds limited only by the specific CCD camera in use \cite{Zhao2019}.
\begin{figure*}[h!]
   \centering
    \includegraphics[width=0.9\textwidth]{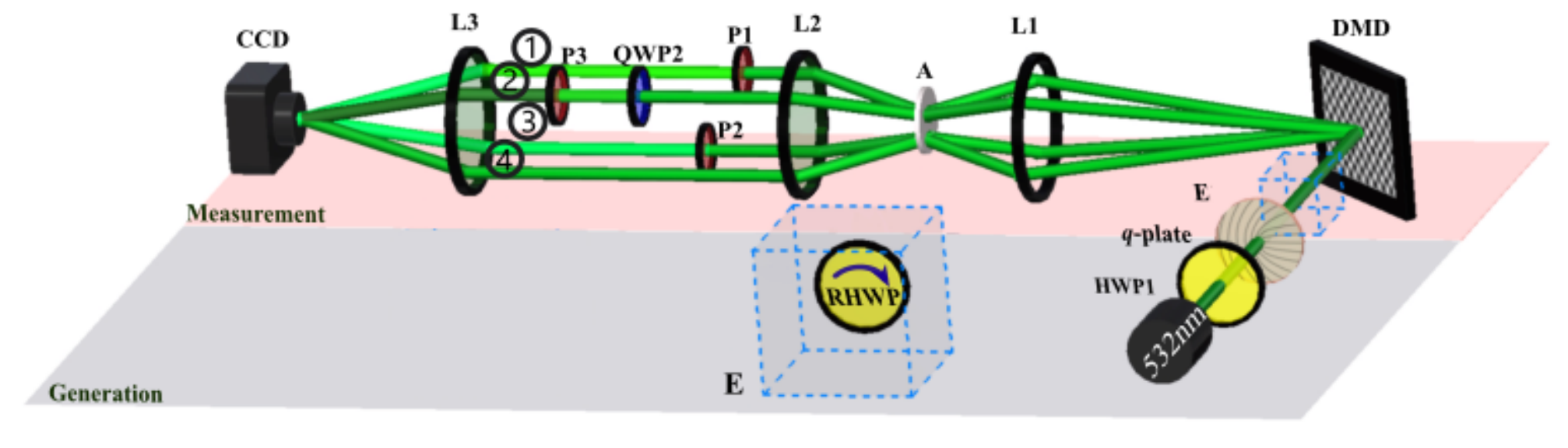}
    \caption{Schematic representation of the setup implemented to reconstruct in real time the SoP of a light field. A CW Gaussian beam ($\lambda=532$ nm) is transformed into a vector beam by the use of a q-plate (q=1/2) and a Half Wave-Plate (HWP1). The resulting beam is then split into four identical copies propagating along parallel paths using a DMD with the help of lenses $L_1$ and $L_2$. $I_H$ and $I_D$ are measured by inserting each path with the linear polarizers P1 and P2, respectively, while P3 in combination with a Quarter Wave-Plate (QWP2) filters $I_R$. Lens $L_3$ $f_3=200$ mm finally focuses these four beams into a CCD to capture all the intensities in one single shot. The system E composed by a Rotating Half Wave-Plate (RHWP) inserted before the DMD, enables real-time evolution of the input beam's SoP. }
   \label{Setuprealtime}
\end{figure*}
A schematic representation of the experimental setup implemented to demonstrate this idea is shown in Fig.\ref{Setuprealtime}, which is divided into two sections: $\mathbf{Generation}$ and $\mathbf{Measurement}$. The required CV beam is generated from a linearly polarized Gaussian beam ($\lambda=532$nm) via a q-plate (q=1/2) in combination with a Half Wave-Plate (HWP1). A multiplex digital hologram with unique diffraction gratings is displayed on the DMD, to split the input beam into four identical copies, as illustrated in Fig. \ref{resultrealtime}(a), from which, the polarisation distribution can be reconstructed. To remove the undesired diffraction orders, these beams are spatially filtered by four individual apertures and collimated to propagate parallel to each other using the set of lenses  $L_1$ and $L_2$ ($f_{1,2}=200$ mm). Afterwards, the required intensities $I_0$, $I_H$, $I_D$ and $I_R$ are measured as explained next (see Fig.\ref{resultrealtime}(b) ). The intensity $I_H$ is measured by inserting a linear polariser at $\theta=0^\circ$ (P1) from  path \encircle{1}. The intensity $I_D$ is obtained by using another linear polariser at $\theta=45^\circ$ (P2) in path \encircle{3}. From  path \encircle{2}, $I_R$ is measured by combining a QWP (QWP2) at $45^\circ$ and a linear polariser at $90^\circ$. Finally, the intensity $I_0$ is directly obtained from path  \encircle{4}. To record all the required intensities in a single image, a third lens (L3, $f_3=200$ mm) is added to focus the beams into a CCD camera (BC106N-VIS from Thorlabs).

\begin{figure}[h!]
   \centering
    \includegraphics[width=0.49\textwidth]{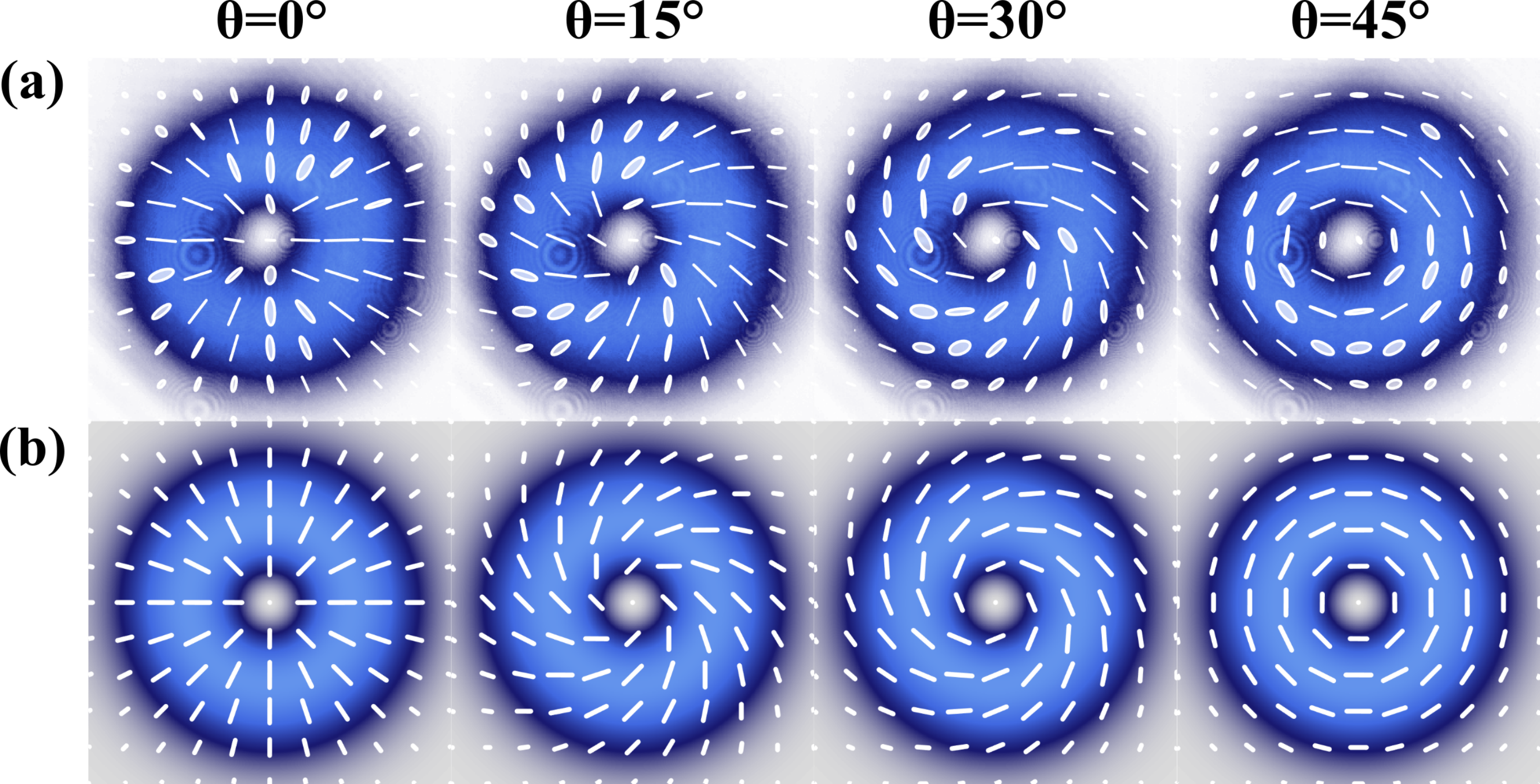}
    \caption{Extracted frames of the real-time reconstruction of polarisation after passing the generated beam through a rotating Half Wave-plate at  $\theta=0^\circ, 16^\circ, 30^\circ$ and $45^\circ$ for
    (a) experiment and (b) simulation.}
   \label{realtimereconstruction}
\end{figure}

The position and intensity of each beam is previously calibrated with a vortex beam, by adjusting the micromirrors in the "On" state, as well as the period of the linear grating, as illustrated in Fig.\ref{resultrealtime}(a). Here, the beams are labelled with the numbers \encircle{1}, \encircle{2}, \encircle{3} and \encircle{4} to identify their position in the experimental setup (Fig.\ref{Setuprealtime}). Figure \ref{resultrealtime}(b) shows an example of the simultaneously measured intensities and the required Stokes parameters reconstructed from these intensities are shown in Fig.\ref{resultrealtime}(c). Finally, the polarisation distribution of the input beam is reconstructed, as detailed in \cite{Goldstein2011}, which for the case at hand yields the radial vector mode shown in Fig.\ref{resultrealtime}(d). This technique allows to monitor in real-time the polarisation evolution of dynamically-changing vector modes, as shown in Fig. \ref{realtimereconstruction}. Here, a Half Wave-Plate, represented as E in Fig. \ref{Setuprealtime}, was inserted in the path of the system to modify in real-time the SoP of the input beam. As the RHWP rotates from $0^\circ$ to $90^\circ$, an angle-dependant phase delay between both orthogonal polarisation components is introduced, resulting in a continuous evolution of the vector modes, from radial (at $0^\circ$ and $90^\circ$) to azimuthal (at $45^\circ$) polarisation. Experimental and numerical simulations are shown in Fig. \ref{realtimereconstruction}(a) and Fig.\ref{realtimereconstruction}(b), respectively, for the specific angle of $0^\circ$, $15^\circ$, $30^\circ$ and $45^\circ$. 

\subsection{Determination of the non-separability through the Vector Quality Factor }

In this section we will describe a technique which allows to quantify the purity of vector modes \cite{McLaren2015,Ndagano2016}.This quantity is based  on the mathematical similarities between classical and quantum entanglement. In quantum mechanics this quantity, known as {\it Concurrence (C)}, measures the degree of entanglement of two photons, in classical optics it is known as Vector Quality Factor (VQF) and also measures the degree of entanglement between the spatial and polarisation degrees of freedom, assigning 1 to vector modes, with a maximal degree of entanglement and 0 to scalar mode, with a null degree of entanglement. In general, the VQF is determined by first projecting the vector mode onto one DoF, either the spatial or polarisation, and tracing over the other by means of spatial or polarisation filters. The first experimental demonstration of this technique was performed with SLMs, for this, a given vector mode is first projected onto the two orthogonal polarisation components, afterwards, the resulting mode is passed through a series of six spatial filters encoded as holograms on the SLM. The VQF is computed from twelve measurements, six for each polarisation components of the on-axis intensity measured in the far field, all of which can be performed simultaneously using a multiplexing approach  \cite{Bhebhe2018a,Ndagano2016,Otte2018}. One of the main drawback in using SLMs is their polarisation dependence, which only allows to modulate linearly polarised states (typically the horizontal), which is the main reason why it is required to first project over the polarisation DoF, by splitting the input vector mode onto the two polarisation components. As such, in this section we will introduce another affordable method to measure the VQF of any arbitrary vector mode by taking full advantage of the polarisation insensitive attribute of DMDs. Noteworthy, this technique enables a reduction in the number of required measurements, from 12 to 8. Additionally, it is low-cost, it can operate over a wide range of wavelengths, and it can measure the VQF in real-time.

\begin{figure}[h!]
   \centering
    \includegraphics[width=0.49\textwidth]{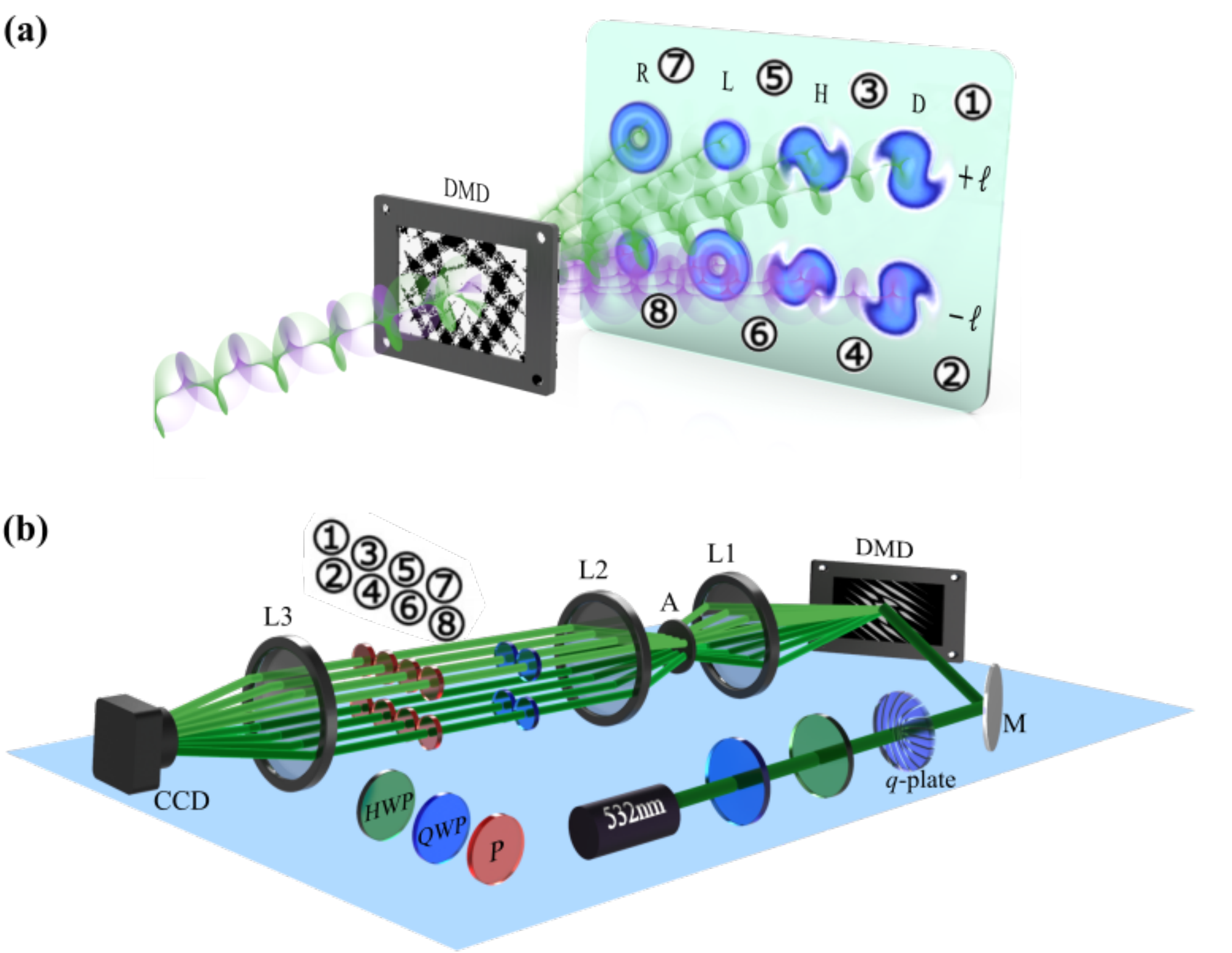}
    \caption{ (a) Conceptual representation of the simultaneous measurement of all the intensities required for computing the VQF using a DMD. (b) Schematic representation of the implemented setup, where CV modes with controllable VQF are generated from a CW Gaussian beam ($\lambda=532$nm) using a liquid crystal q-plate (q=1/2) in combination with a Half- (HWP) and Quarter Wave Plate(QWP1). The multiplexing hologram encoded in the DMD enables the simultaneous projection onto the OAM basis and the resulting modes are then passed through a series of optical filters to trace over the polarisation DoF.  A "single shot" of all the intensities can be recorded using a CCD camera in combination with a lens $L_3$. }
   \label{VQFsetup}
\end{figure}

To begin with, let us remind that the VQF can be computed as the real part of the degree of concurrence {\it C}, as defined by Wooters \cite{Wootters1998}. More explicitly, it has the form \cite{Ndagano2016}, 
\begin{equation}
\text{VQF} =\text{Re}(C)=\text{Re}\left(\sqrt{1-s^2}\right)=|\sin(2\theta)|,
\label{eq:VQF}
\end{equation} 
where $s$ is the length of the Bloch vector defined as,
\begin{equation}
s = \left(\sum_{i=1}^3 \langle\sigma_i\rangle^2\right)^{1/2}.
\label{eq:Blochvector}
\end{equation}
Here, $\langle\sigma_1\rangle$, $\langle\sigma_2\rangle$ and $\langle\sigma_3\rangle$ are the expectation values of the Pauli operators, which represent a set of normalised intensity measurements. For the specific case of cylindrical vector modes, these intensities are obtained by first projecting the vector mode on the $|\pm \ell \rangle$ OAM basis, encoded on the DMD, and tracing afterwards over the polarisation DoF using a series of polarisation filters. Importantly, all intensities can be measured simultaneously using a multiplexing approach, as schematically represented in Fig.~\ref{VQFsetup} (a). To this end, eight individual holograms are multiplexed with unique spatial frequencies to diffract each beam along different angles. As illustrated in this figure, the top four holograms perform the $| +\ell \rangle$ projection, while the four on the bottom perform the $| -\ell \rangle$ projection. Each beam is then passed through a series of polarisation filters to trace over the polarisation DoF, namely, the polarisation components $| R \rangle$, $| L \rangle$, $| H \rangle$ and $| D \rangle$ . At last, a lens performs the far-field of all beams simultaneously and the on-axis intensity of each of them is measured to obtain the expectation values $\langle\sigma_1\rangle$, $\langle\sigma_2\rangle$ and $\langle\sigma_3\rangle$ (see \cite{Zhaobo2020} for more details). The required projections are shown in Table \ref{Tomographic}. Here, for example, $I_{R\ell^-}$ represents the intensity after projecting the vector mode on the $| -\ell \rangle$ OAM phase filter and passing it through a $| R \rangle$ polarisation filter. 
\begin{table}[ht!] 
\setlength{\tabcolsep}{10pt}
\renewcommand*{\arraystretch}{2}
 \caption{Normalised intensity measurements $I_{mn}$ to determine the expectation values $\langle \sigma_i \rangle$. \label{Tomographic}}
 \begin{tabular}{c|c c c c c }
Basis states & $| R \rangle$&$| L \rangle$&$| H \rangle$&$| D \rangle$&\\ 
\hline \hline
$| +\ell \rangle$& $I_{R\ell^+}$&$I_{L \ell^+}$&$I_{H\ell^+}$&$I_{D \ell^+}$&\\
$| -\ell \rangle$& $I_{R\ell^-}$ & $I_{L \ell^-}$ & $I_{H\ell^-}$ & $I_{D \ell^-}$
 \end{tabular}
 \end{table}
 
Using the eight intensities shown in Table \ref{Tomographic}, the expectation values $\langle\sigma_i\rangle$ take the explicit form, 
\begin{align}
\centering
\nonumber
   &\left \langle \sigma_1 \right \rangle=2(I_{H \ell^+}+I_{H \ell^-})-(I_{\ell^+}+I_{\ell^-}),\\
   &\left \langle \sigma_2 \right \rangle=2(I_{D \ell^+}+I_{D \ell^-})-(I_{\ell^+}+I_{\ell^-}),\label{eq:Pauli}\\
    \nonumber
    &\left \langle \sigma_3 \right \rangle=2(I_{R \ell^+}+I_{R \ell^-})-(I_{\ell^+}+I_{\ell^-}),
\end{align}
\noindent
where,  $I_{\ell^+}=I_{R \ell^+}+I_{L \ell^+}$ and $I_{\ell^-}=I_{R \ell^-}+I_{L \ell^-}$.

Figure \ref{VQFsetup}(b) shows a schematic representation of the implemented experimental setup. Here, the input vector mode is first sent to the center of DMD where the eight holograms are displayed to perform the projection onto the spatial DoF. As can be seen, after the DMD eight beams emerge along different angles, which are passed afterwards through a 4$f$ system consisting of lenses $L_1$ and  $L_2$ ($f_{1,2}$=200mm). An aperture (A) placed in the focusing plane of  $L_1$ allows to filter only the first diffraction order of each beam. After performing the projection on spatial basis $| +\ell \rangle$ (beams \encircle{1},\encircle{3},\encircle{5},\encircle{7}) and  $| -\ell \rangle$ (beams \encircle{2},\encircle{4},\encircle{6},\encircle{8}), the next step is to trace over the polarisation DoF by inserting to each path specific different polarisation filters. To be more specific, $I_{H \ell^+}$ and $I_{H \ell^-}$ are obtained by passing beams  \encircle{1} and  \encircle{2} through a linear polariser(P) at $0^\circ$,   $I_{D \ell^+}$ and $I_{D \ell^-}$ are acquired by passing beams  \encircle{3} and  \encircle{4} through a linear polariser oriented at  $45^\circ$. To measure the  $I_{R \ell^+}$ and $I_{R \ell^-}$, a Quarter Wave-Plate (QWP) at  $45^\circ$ in combination with a linear polariser at  $90^\circ$ are inserting into the paths \encircle{5} and \encircle{6}. Finaly, $I_{L \ell^+}$ and $I_{L \ell^-}$ are acquired by transmitting beams \encircle{7} and \encircle{8} through a QWP at $-45^\circ$ and a linear polariser at $90^\circ$. To measure the far field intensities in a "single shot", all beams are focused onto a CCD camera (BC106N-VIS Thorlabs) with the help of a lens ($L_3$,$f_3$=200mm).

\begin{figure}[tb]
   \centering
    \includegraphics[width=0.39\textwidth]{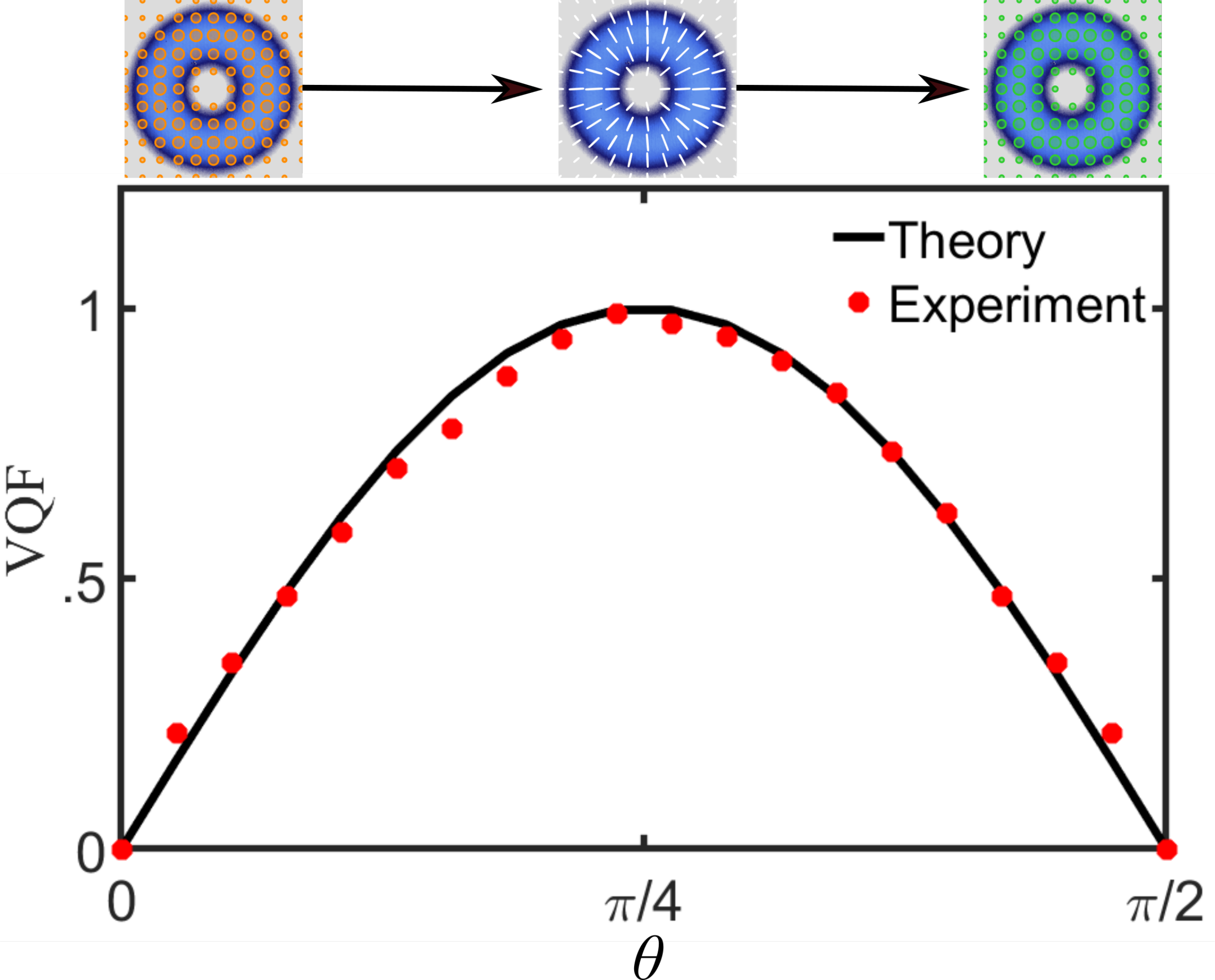}
    \caption{VQF as function of the parameter $\theta$. The black solid line corresponds to the theory and the red data points to the experimental results. Here, we show the CV modes with continuous varying inter-modal phase $\theta\in[0,\pi/2]$. Notice the entanglement reaches its maximum for pure vector mode $\theta=\pi/4$ while it gains null for scalar mode $\theta=0, \pi/2$. The insets show the polarisation distribution of such three typical cases, where the colors of orange and green represent the right and left circular polarisation while the white stands for the linear. }
   \label{VQFresult}
\end{figure}

Figure.\ref{VQFresult} shows experimental results of the VQF as a function of $\theta$, where the theory (Eq.\ref{eq:VQF}) is represented by the black continuous line and the experimental data  by red points. Here, we show the VQF increasing continuously from 0 to 1 and back to 0, corresponding to the scalar ($\theta=0$), pure vector ($\theta=\pi/4$) and scalar ($\theta=\pi/2$) modes, respectively. The corresponding polarisation distribution for these three typical cases are shown in the insets of Fig.\ref{VQFresult}, where the orange and green colors correspond to the right and left polarisation states, respectively, and the white color to the linear polarisation. Notice the high agreement of the experimental data with the theoretical prediction. 

Crucially in recent time it was also proposed that the degree of coupling between the spatial and polarisation degrees of freedom can be measured directly from the Stokes parameters, by integrating these over the transverse plane \cite{selyem2019}. In this way, no priory knowledge of the involved spatial modes is needed for the reason that these polarisation measurements for obtaining the required Stoke parameters are nevertheless affected by the spatial structure of the vector beam. More specifically, the degree of coupling can be mathematically expressed as $\sqrt{1-(\mathbb{S}_1^2+\mathbb{S}_2^2+\mathbb{S}_3^2)/\mathbb{S}_0^2}$, where $\mathbb{S}_i$ are the values of the Stokes parameters $S_i$ integrated over the entire transverse profiles. This approach offers a basis-independent way to infer the degree of coupling between both DoFs by using only the conventional Stokes parameters, representing a notable advance in characterizing these complex vector light fields.

\section{Conclusions}
The techniques presented in this tutorial provides a framework for the generation and characterisation of vector modes employing a Digital Micromirror Device.  Even though SLMs have already shown their capabilities in shaping complex light fields, the  unnoticed attribute of DMDs offers more flexibility and exhibits competitive advantage.  Here, we firstly introduce  a novel technique allowing the generation of arbitrary vector mode by fully taking advantage of the polarisation-insensitive attribute of DMDs. Together with the employment of spatial random multiplexing encoding scheme, this technique enables to thoroughly exploit the rather high refresh rate of DMDs during the generation process. We further demonstrated this technique not only by experimentally generating the well-known CV mode, but also by generating another three types of vector modes whose spatial basis are encoded with the Ince-,  Mathieu-, and Parabolic-Gauss beams. The high agreement between the experimental results and theoretical prediction exhibit the reliable performance of this technique. In terms of the characterisation of vector modes, we explained another two novel DMD-techniques capable to monitor in real time complex vector light field, one through the SP allowing to reconstruct the polarisation distribution of the generated beams; the other one through the VQF which provide a reliable way to  determine in real-time the non-separability of the undetected beams. All the work present here provides new light on the field of structured light by employing DMDs with high refresh rates allowing advanced applications in fields such as optical communications, optical metrology and optical tweezers, to mention a few.

\section*{Acknowledgement} 

\section*{Funding}
This work was partially supported by the National Natural Science Foundation of China (NSFC) under Grant No.  61975047.

\section*{Disclosures}
The authors declare that there are no conflicts of interest related to this article

\section*{References}
\bibliographystyle{iopart-num}
\bibliography{References}

\end{document}